\documentclass[a4paper,11pt]{article}
\pdfoutput=1 
\usepackage{jheppub} 
\usepackage[T1]{fontenc} 
\usepackage{multirow}
\RequirePackage[T1]{fontenc}

\RequirePackage{graphicx}
\RequirePackage{mathptmx} 
\RequirePackage{flushend}
\RequirePackage[numbers,sort&compress]{natbib}
\usepackage{subfigure}
\makeatletter
\def\BibitemShut#1{\unskip} 
\makeatother
\usepackage[normalem]{ulem}
\usepackage{graphicx}
\usepackage{amsmath,amssymb}
\usepackage{dcolumn}
\usepackage{bm}
\usepackage{color}
\usepackage{appendix}
\newcommand{\f}{\frac}
\newcommand{\lt}{\left}
\newcommand{\n}{\nonumber}

\newcommand{\rt}{\right}
\newcommand{\rr}{\mathrm}
\newcommand{\dd}{{\mathrm d}}

\newcommand{\pbh}{\mathrm{PBH}}

\title{\boldmath Memory burden effect of regular primordial black holes}
\author[a]{Jin-Rong Du,} 
\author[a]{Zi-Zhuo Zhang,}
\author[a]{Nan Li}
\affiliation[a]{Department of Physics, College of Sciences, Northeastern University, Shenyang, 110819, China}
\emailAdd{dujinrong61@gmail.com}
\emailAdd{zzizhuo123@163.com}
\emailAdd{linan@mail.neu.edu.cn}

\abstract{Primordial black holes (PBHs) have attracted intensive research interest as a promising candidate of dark matter. However, because of the Hawking radiation, the PBHs lighter than $10^{15}~\rm{g}$ have already evaporated before today. To extend the PBH mass window to small-mass range, two possible ingredients are explored. The first is the consideration of regular PBHs with non-singular metrics, which can decrease the Hawking temperature, thereby lowering black hole evaporation. The second is the incorporation of the memory burden (MB) effect, which can further suppress the evaporation rate, after regular PBHs have lost a certain amount of their initial masses. In this work, we combine these two ingredients and study the MB effects of three types of regular PBHs (the Hayward, Bardeen and Simpson--Visser black holes). Assuming a phenomenological self-similar evaporation, we find that the MB effect significantly relaxes the evaporation constraints. For a benchmark of the MB strength parameter $k=1$, a new PBH mass window opens at around $10^6$--$10^8$ g, where regular PBHs can compose all dark matter without violating the Big Bang nucleosynthesis bounds.}

\keywords{Regular black holes, Primordial black holes, Memory burden, Dark matters}

\begin{document}
\maketitle
\flushbottom

\section{Introduction} \label{sec:intro}

Dark matter (DM) is one of the central unresolved puzzles in modern cosmology. Astronomical observations have provided various evidence for its existence, including galaxy rotation curves, gravitational lensing, and the formation of large-scale structures. Current measurements indicate that DM approximately constitutes $27\%$ of the total energy density in the present Universe \cite{Planck:2018vyg}; however, its particle nature has yet to be determined. Although the standard cosmological model has achieved remarkable successes in describing the evolution of the Universe, the precise composition of DM remains an open question.

Traditionally, the weakly interacting massive particles (WIMPs) have been considered as a prominent candidate of DM, usually motivated by theories beyond the Standard Model of particle physics \cite{Arcadi:2017kky, Schumann:2019eaa}. Nevertheless, after decades of efforts, no conclusive evidence for WIMPs has been confirmed in direct detections \cite{SuperCDMS:2017mbc, LZ:2024zvo, XENON:2025vwd}, indirect searches \cite{Klasen:2015uma, Conrad:2017pms}, or collider experiments \cite{Escudero:2016gzx, CMS:2017jdm}. This situation has motivated the explorations of non-particle DM candidates, including axions \cite{Marsh:2015xka, Salvio:2021puw, Berlin:2016bdv}, sterile neutrinos \cite{Boyarsky:2018tvu, Drewes:2016upu, Abazajian:2017tcc}, and, the focus of this work, primordial black holes (PBHs) \cite{Carr:1974nx, Carr:1975qj, Carr:2005zd, Carr:2009jm, Carr:2020gox, Carr:2020xqk, Auffinger:2022khh, Green:2024bam}.

Unlike astrophysical black holes formed from stellar collapse at late times, PBHs can originate from large density perturbations in the very early Universe. When the local density contrast exceeds a certain threshold, gravitational collapse can directly produce a PBH. As a result, PBHs can span an extremely wide mass range, from the Planck mass up to $10^3~M_\odot$ ($M_\odot=1.99 \times 10^{33}~\rm{g}$ is the solar mass). At present, the scenario that PBHs constitute the dominant component of DM is facing challenges. According to evaporation, lensing, dynamics, accretion, and gravitational-wave constraints, there remains only one mass window at $10^{-17}$--$10^{-13}~M_\odot$ (i.e., from the asteroid to the sublunar mass range), where PBHs are possible to compose all DM \cite{Carr:2016drx, Thoss:2024hsr}. Moreover, the very light PBHs with masses less than $10^{15}$~g (i.e., $10^{-18}~M_\odot$) are generally thought to have evaporated, and cannot serve as a viable candidate of DM today \cite{Green:2024bam, Montero-Camacho:2019jte}.

Consequently, how to extend the allowed mass range for PBHs, especially in the small-mass direction, is receiving research interest recently. In particular, the proposal of the memory burden (MB) effect \cite{Dvali:2018xpy, Dvali:2018ytn, Dvali:2020wft, Dvali:2024hsb} aims to open a potential mass window below $10^{23}$~g. In Ref. \cite{Dvali:2024hsb}, the authors first identified a swift MB effect, according to which black holes can experience an effective MB that grows as evaporation proceeds and progressively suppresses the evaporation rate. An immediate implication is that small-mass black holes can acquire significantly prolonged lifetimes, allowing them to survive through the constraints from Big Bang nucleosynthesis (BBN) and cosmic microwave background (CMB), thereby reopening the previously excluded small-mass window for PBHs \cite{Alexandre:2024nuo, Chianese:2024rsn, Zantedeschi:2024ram}. Within this framework, there has been a surge of studies on small-mass PBHs \cite{Balaji:2024hpu, Jiang:2024aju, Haque:2024eyh, Basumatary:2024uwo}, gravitational waves \cite{Barman:2024iht, Loc:2024qbz, Kohri:2024qpd, Bhaumik:2024qzd, Athron:2024fcj, Barker:2024mpz, Maity:2024cpq}, and other physical processes associated with the PBH formation and evaporation \cite{Sarmah:2025wrg, Boccia:2025hpm, Anchordoqui:2024tdj, Bandyopadhyay:2025ast, Borah:2024bcr, Domenech:2024wao}.

Subsequently, Refs. \cite{Montefalcone:2025akm, Dvali:2025ktz} independently pointed out that, in the MB scenario, small-mass black holes should be subject to the evaporation-based constraints similar to those applicable to large-mass black holes. In Ref. \cite{Montefalcone:2025akm}, the authors further demonstrated that the swift MB effect can significantly strengthen evaporation constraints, rendering the small-mass window nearly closed. In response, Ref. \cite{Dvali:2025ktz} proposed a modified slow MB effect, in which black holes rapidly enter a strong MB region, thereby weakening evaporation constraints and effectively resolving the overly restrictive bounds from the swift MB effect. More recently, comparative studies on the swift and slow MB effects, as well as further related developments and criticisms, have been carried out \cite{Chaudhuri:2025rcs, Chaudhuri:2025asm, Tan:2025vxp, Dondarini:2025ktz, Ettengruber:2025kzw, Maity:2025ffa, Tseng:2025fjf, Yuan:2025hls}. In particular, the MB effect of PBHs can be found in Refs. \cite{Dvali:2020wft, Alexandre:2024nuo, Anchordoqui:2024dxu, Balaji:2024hpu, Dvali:2024hsb, Haque:2024eyh, Barman:2024iht, Bhaumik:2024qzd, Barker:2024mpz, Jiang:2024aju, Anchordoqui:2024tdj, Bandyopadhyay:2025ast, Boccia:2025hpm, Loc:2024qbz, Borah:2024bcr, Maity:2024cpq, Basumatary:2024uwo, Domenech:2024wao, Chianese:2024rsn, Kohri:2024qpd, Montefalcone:2025akm, Dvali:2025ktz, Calabrese:2025sfh, Athron:2024fcj, Chaudhuri:2025rcs, Tan:2025vxp, Dondarini:2025ktz, Chaudhuri:2025asm, Ettengruber:2025kzw, Maity:2025ffa, Chianese:2025wrk, Zantedeschi:2024ram, Gross:2025hia, Perez-Gonzalez:2025try, Dvali:2025sog, Merchand:2025bzt, Tseng:2025fjf, Yuan:2025hls, Levy:2025lyj, Kitabayashi:2025iaq, Sarmah:2025wrg, Leontaris:2025piz}.


We notice that the MB effect opens a mass window for PBHs by suppressing the evaporation rate. In current studies, the relevant discussions on PBHs are focused on singular black hole solutions (e.g., the Schwarzschild and Kerr metrics), but we find that incorporating regular black holes (RBHs) can similarly decrease the black hole evaporation rate. Therefore, in this work, we aim to investigate the MB effect of regular PBHs. Notable examples of RBHs include phenomenological models \cite{Bronnikov:2000vy, Ayon-Beato:1999kuh, Antoniou:2017acq, Bambi:2013ufa, Fan:2016hvf, Chong:2005hr, Bronnikov:2005gm, Abdujabbarov:2016hnw, Balart:2014cga, Johannsen:2013szh, Johannsen:2011dh, Boonserm:2018orb}, such as the Hayward metric \cite{Hayward:2005gi, Hayward:1993wb}, Bardeen metric \cite{Ayon-Beato:1998hmi, Ayon-Beato:2000mjt,Bardeen:1968}, and Simpson--Visser metric \cite{Simpson:2018tsi, Simpson:2019mud}. Although certain RBH models remain constrained by the issues like geodesic incompleteness and challenges regarding their dynamical formation \cite{Bueno:2025zaj}, these metrics nonetheless serve as foundational models for phenomenological explorations. In light of this, it is of particular interest and importance to investigate such regular PBHs and to evaluate their potential in elucidating the constituents of DM \cite{Calza:2024fzo, Calza:2024xdh, Calza:2025mwn, Davies:2024ysj, Asmanoglu:2025agc, Loc:2025mzc}.

Since both the MB effect and RBHs can weaken the black hole evaporation, lengthen the black hole lifetime, and thus open new small-mass window for PBHs, the aim of our present work is a comprehensive study on the MB effect of regular PBHs. The key novelty is the incorporation of the MB effect to explore and constrain the parameter spaces of regular PBHs as a candidate of DM. We study the evaporation of RBHs by employing a modified Hawking temperature formula.\footnote{It is important to note that, although the rigorous dynamical formation of RBHs typically necessitates higher-dimensional spacetime, we treat them herein as effective toy models within a four-dimensional framework for phenomenological discussions, without delving into the specific details of their formation processes.} Our results indicate that, compared to singular PBHs, regular PBHs influenced by the MB effect can accommodate a much broader mass range, capable of accounting for all DM in the small-mass range at around $10^6$--$10^8$~g. While this conclusion is phenomenological in nature, it compellingly demonstrates the significance and potential of non-singular black hole geometries in the realm of PBH cosmology.

This paper is organized as follows. The MB effect is reviewed in Sec. \ref{sec:MB}. Then, the physical properties of three types of RBHs (i.e., the Hayward, Bardeen, and Simpson--Visser black holes) are presented in Sec. \ref{sec:regular}. In Sec. \ref{sec:dis}, we discuss the evaporation of regular PBHs, and find that the MB effect can significantly relax the BBN constraints. Furthermore, a detailed discussion is provided on the sensitivity of different metrics to the value of $k$. We conclude in Sec. \ref{sec:con}. We work in the natural system of units and set $c=G_{\rm N}=\hbar=k_{\rm B}=1$.

\section{MB effect} \label{sec:MB}

In this section, we briefly review the core principles of the MB effect and describe its role in our following analysis. We adopt the swift MB effect as an example \cite{Dvali:2024hsb, Haque:2024eyh}, but our main conclusions also apply to the slow MB regime.

In black hole evaporation, the MB effect originates from the dependence of black hole evolution on the information associated with its formation history and internal degrees of freedom \cite{Dvali:2018xpy, Dvali:2018ytn, Dvali:2020wft, Dvali:2024hsb}. Here, “memory” refers to the microscopic information encoded in the black hole interior or its quantum state, such as the initial formation conditions.
This information is not instantaneously or completely erased by the Hawking radiation; rather, it continues to influence the quantum evolution of the black hole.
Hence, “burden” refers to the dynamical cost imposed by the retention of this information in the evaporation process. As the black hole mass decreases, the relative energy or phase-space cost required to sustain these internal degrees of freedom increases, significantly suppressing the Hawking radiation rate and effectively delaying the black hole evaporation.


We start from the standard black hole thermodynamics \cite{Hawking:1974rv}, and describe the semi-classical evaporation rate of a black hole as
\begin{align} 
\lt.\frac{\dd M}{\dd t}\rt|_{\rm{sc}}= -4\pi R^2\sigma T^4, \label{SCevap}
\end{align}
where $M$ and $R$ are the black hole mass and outer horizon radius, $T$ is the Hawking temperature, and $\sigma$ is the effective Stefan--Boltzmann constant accounting for all the species of particles emitted during the Hawking radiation. Equation (\ref{SCevap}) provides the basis for calculating the mass-loss rate due to the Hawking radiation in the semi-classical regime. 

In Ref. \cite{Dvali:2024hsb}, the authors suggested that the black hole evaporation rate can be suppressed by its entropy in the MB effect. Under the circumstances, Eq. (\ref{SCevap}) is modified as
\begin{align}
\lt.\lt.\frac{\dd M}{\dd t}\rt|_{\rm{MB}} =\frac{1}{S^{k}}\frac{\dd M}{\dd t}\rt|_{\rm{sc}}, \label{MBF}
\end{align}
where $S$ is the black hole entropy, and the semi-classical evaporation rate in Eq. (\ref{SCevap}) is suppressed by a factor $S^{-k}$, with $k$ being a parameter describing the strength of the MB effect, typically chosen as a positive integer (for the discussion on the $k$-dependence of the MB effect, see Sec. \ref{sec:k}).

Furthermore, a characteristic feature of the swift MB effect is that the black hole first undergoes a period of semi-classical evaporation, and then enters the MB regime only after losing a certain amount of its initial mass, at which point the evaporation rate transitions from Eq. (\ref{SCevap}) to Eq. (\ref{MBF}). The black hole mass at this transition is usually written as
\begin{align}
M_{q} =q M_{\rm ini}, \n
\end{align}
and the timescale to reach this stage can be obtained as \cite{Haque:2024eyh}
\begin{align}
t_{\rm{sc}}=(1-q^3)\tau_{\rm{sc}}, \label{ttMB}
\end{align}
where $M_{\rm ini}$ denotes the initial mass of the black hole at the onset of evaporation, and the parameter $q$ indicates the transition from the semi-classical evaporation to the swift MB-dominated regime. Throughout this work, we follow Ref. \cite{Dvali:2024hsb}, and adopt the value as $q =1/2$. Moreover, the quantity $\tau_{\rm sc}$ represents the semi-classical evaporation lifetime of the black hole (e.g., $\tau_{\rm sc}=5120\pi M_{\rm ini}^3$ for the Schwarzschild black hole \cite{Hawking:1975vcx}).

Above, we have reviewed the role of the MB effect in delaying the black hole evaporation. In the following sections, we will further discuss its implications on regular PBHs in detail.

\section{Regular black holes} \label{sec:regular}

In this section, we introduce three distinct RBH metrics and present two different prescriptions for selecting the regularization parameter. Then, we discuss the modified Hawking temperature and reformulate the constraints on black hole evaporation.

Assuming that the metric description remains valid, we introduce a metric that is non-singular throughout the entire spacetime. Considering the $t$--$r$-symmetric black hole metric, the line element can be expressed as
\begin{align}
\dd s^2=-F(r)\,\dd t^2+\f{\dd r^2}{F(r)} +G(r)\,\dd \Omega ^2, \n
\end{align}
where $F(r)$ and $G(r)$ are the metric functions, and ${\rm d}\Omega$ is the solid angle. We require the metric functions to satisfy the condition of asymptotic flatness, which implies that
\begin{align}
F(r)\to 1, \quad G(r)\to r^2 \qquad (r\to\infty). \n
\end{align}

Below, we explore three typical RBH metrics: the Hayward, Bardeen, and Simpson–Visser metrics, respectively. 

\subsection{Hayward black hole} \label{sec:Hayward}

The metric functions in the Hayward spacetime read
\begin{align}
F_{\rr H}(r)=1-\f{2 M r^2}{r^3 +2M \ell^2}, \quad G_{\rm H} (r)=r^2, \label{Hayward}
\end{align}
where the positive regularization parameter $\ell$ serves as a fundamental length scale that ensures the resolution of the singularity in the limit $r\to 0$. Naturally, when $\ell \to 0$, the Hayward metric reduces to the Schwarzschild metric. 

If we require the metric functions to describe a black hole solution, the constraint $\ell\leq\ell_*= 4M/(3\sqrt{3})$ must be satisfied. Therefore, assuming the metric describes a black hole throughout its evolution, we obtain a critical mass $M_* =3\sqrt{3}\ell/4$ and a corresponding critical horizon radius $R_* = \sqrt{3}\ell$. Beyond this critical limit, the horizon vanishes. 

Below, we consider two distinct models for the evaporation of RBHs.
\begin{enumerate}
\item The parameter $\ell$ is treated as a constant, implying that the black hole cannot evaporate completely. Due to this constant $\ell$, the evaporation process differs from the Schwarzschild case and loses self-similarity.

\item The parameter $\ell$ is assumed to be proportional to the horizon radius $R$, and the self-similarity of the black hole evaporation is preserved.
\end{enumerate}

\begin{figure}[htb]
\centering \includegraphics[width=0.6\linewidth]{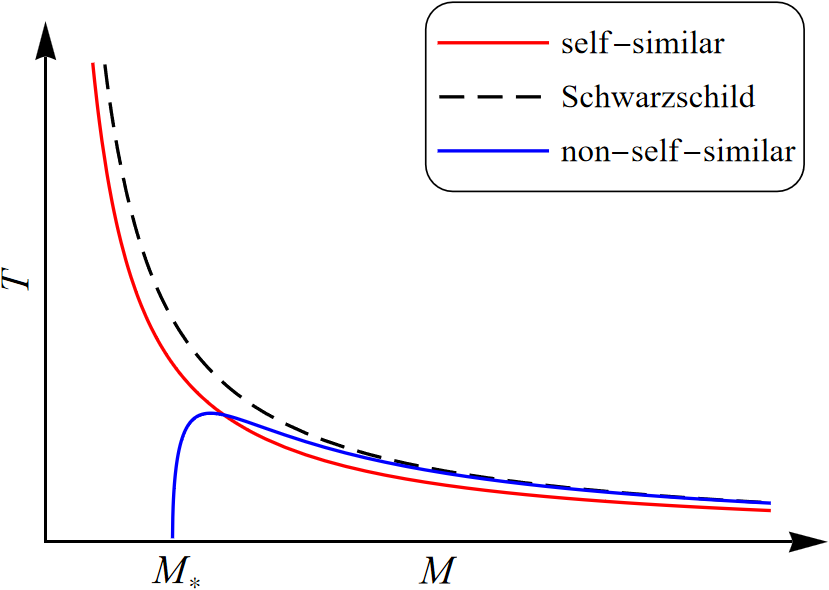}
\caption{The Hawking temperature $T$ as a function of the Hayward black hole mass $M$. The dashed black line corresponds to the Schwarzschild black hole. The solid lines represent the RBH solutions under different regularization schemes: the red line denotes the self-similar case, while the blue line is for the non-self-similar case, respectively. Notably, the non-self-similar evaporation exhibits a maximum temperature, and then the temperature tends to zero as $M$ approaches its critical value $M_*$, indicating the formation of a cold remnant.} \label{fig:BHTM}
\end{figure}

Figure \ref{fig:BHTM} illustrates the thermal evolution of the Hayward black holes, depicting the Hawking temperature $T$ as a function of black hole mass $M$, for the Schwarzschild case (black dashed line), self-similar case (red line), and non-self-similar case (blue line), respectively. In the asymptotic large-mass region ($M \gg \ell$), all three curves converge. This indicates that for macroscopic black holes, the regularization effects become negligible, thereby recovering the standard Schwarzschild behavior ($T \propto 1/M$). However, significant discrepancies emerge in the final stages of evaporation (the small-mass region). The temperature of the Schwarzschild black hole diverges as its mass vanishes, leading to the catastrophic endpoint of evaporation. The self-similar evaporation is basically similar to the Schwarzschild profile, only with its temperature slightly lower. In contrast, the non-self-similar evaporation displays a notably distinct feature: the black hole temperature initially increases to a peak, and then decreases to zero as its mass $M$ approaches the critical value $M_\ast$. This suggests that the black hole may evolve into a stable zero-temperature remnant, avoiding the terminal singularity. However, the time required is effectively infinite, so the formation of such remnants is precluded within the finite age of the Universe. Therefore, we will focus our investigation on the self-similar scenario, which represents a more physical framework. Nevertheless, this self-similar evaporation model is essentially a hypothetical toy model, not derived from an underlying theory of RBH formation or evaporation, so our conclusions in this work should be considered model-dependent.

In the self-similar evaporation, where $\ell$ is proportional to the horizon radius, the evaporation rate slows down but does not completely cease, and no remnants are produced. Specifically, we consider a model with the relation as
\begin{align}
\ell = b R_{\rm H} \qquad (0<b<1). \n
\end{align}
The primary advantage of our model is that the self-similarity of the black hole evaporation can be preserved. In this framework, the metric function in Eq. (\ref{Hayward}) becomes
\begin{align}
F_{\rr H}(r)=1-\frac{2M r^2}{r^3+2M(b R_{\rr H})^2}, \label{Hmetric}
\end{align}
and the corresponding horizon radius $R_{\rr H}$ can be obtained as
\begin{align}
R_{\rr H}=2(1- b ^2)M. \n
\end{align}

Last, we turn to the Hawking temperature of the Hayward black hole, $T_{\rr H}={F'_{\rm H}(R_{\rr H})}/({4\pi})$, where $'$ denotes the derivative with respect to $r$.
From Eq. (\ref{Hmetric}), we obtain
\begin{align}
T_{\rr{H}}=\frac{1-3 b^2}{8\pi(1- b^2)M}. \n
\end{align}
Then, from the thermodynamic relation $\dd S=\dd M/T$, the entropy of the Hayward black hole reads
\begin{align}
S_{\rr H}=\frac{4\pi(1-b^2)M^2}{1-3b^2}. \n
\end{align}
To guarantee the positive definiteness of $S$, we obtain the range for the parameter $b$ as $b< 1/\sqrt{3}$.

\subsection{Bardeen black hole} \label{sec:Bardeen}

In 1968, Bardeen proposed a pioneering model of RBH \cite{Bardeen:1968}. By incorporating non-linear electrodynamics as the matter source, this model effectively prevents the formation of a singularity without violating the weak energy condition. Physically speaking, this mechanism can be interpreted as the presence of a magnetic charge inducing the non-linear effects that counteract the gravitational collapse. 

The metric functions for the Bardeen black hole read
\begin{align}
F_{\rr B}(r)=1-\frac{2 Mr^2}{(r^2+\ell ^2)^{3/2}}, \quad G_{\rr B}(r)=r^2. \n
\end{align}
To ensure that this metric describes a black hole solution, the parameter $\ell$ must satisfy $\ell \leq \ell_\ast= 4M/(3\sqrt{3})$ (the same as that for the Hayward black hole). As stated above, we focus exclusively on the self-similar evaporation, so we adopt the regularization parameter as $\ell = bR_{\rm B}$.

Since the calculation details are analogous to the Hayward case, we directly present the relevant results. First, the horizon radius of the Bardeen black hole is
\begin{align}
R_{\rr{B}}=\f{2 M}{(1+b^2)^{3/2}}. \n
\end{align}
The corresponding Hawking temperature is given by
\begin{align}
T_{\rr{B}}=\frac{(1-2 b^2)\sqrt{1+b^2}}{8\pi M}, \n
\end{align}
and the entropy of the Bardeen black hole is obtained as
\begin{align}
S_{\rr {B}}=\frac{4\pi M^2}{(1-2b^2)\sqrt{1+b^2}}. \n
\end{align}
It is easy to find the range for $b$ as $b< 1/\sqrt{2}$.

\subsection{Simpson--Visser black hole} \label{sec:SV}

In RBH thermodynamics, the model proposed by Simpson and Visser stands out due to its unique internal structure \cite{Simpson:2019mud}. Unlike the Hayward and Bardeen metrics, which possess the de Sitter cores characterized by finite curvature and constant energy density at the origin, the Simpson--Visser model features a black bounce throat. The center is not a point but a minimal surface of radius $\ell$, connecting two asymptotically flat universes. Falling matter passes through this throat and bounces into another region. This implies that the spacetime curvature attains a finite maximum as $r \to 0$ (i.e., the spacetime is highly curved in this region). From a phenomenological point of view, the Simpson--Visser black hole introduces distinct corrections to the Hawking temperature and evaporation rate compared to the Hayward and Bardeen black holes. 

The metric functions for the Simpson--Visser black hole are given by
\begin{align}
F_{\rr {SV}}(r)=1-\frac{2M}{\sqrt{r^2+\ell ^2}}, \quad G_{\rr {SV}}(r)=r^2+\ell^2. \n
\end{align}
Different from the Hayward and Bardeen metrics, the derivative satisfies $F'_{\rr {SV}}(r)\ge0$ 
throughout the domain $r\ge 0$. To ensure that this geometry describes a black hole solution, we must require $F_{\rr {SV}}(r)\le 0$ as $r\to 0$, which imposes a constraint on the regularization parameter as $\ell \le 2M$. We notice that the previously employed ansatz $\ell =bR_{\rm SV}$ yields no further constraint on $b$ in this specific model. Therefore, to maintain a parametrization strategy consistent with our previous analysis, we explicitly set $\ell =2bM$ ($0<b<1$).

Again, omitting the intermediate calculations, we obtain the horizon radius of the Simpson--Visser black hole as
\begin{align}
R_{\rr{SV}}=2\sqrt{1-b^2}M. \n
\end{align}
Similarly, the corresponding Hawking temperature reads
\begin{align}
T_{\rr{SV}}=\frac{\sqrt{1-b^2}}{8\pi M}, \n
\end{align}
and the entropy of the Simpson--Visser black hole is
\begin{align}
S_{\rr {SV}}=\f{4\pi M^2}{\sqrt{1-b^2}}. \n
\end{align}
It is easy to see the range for $b$ as $b<1$.

\begin{figure}[htb]
\centering \includegraphics[width=0.65\linewidth]{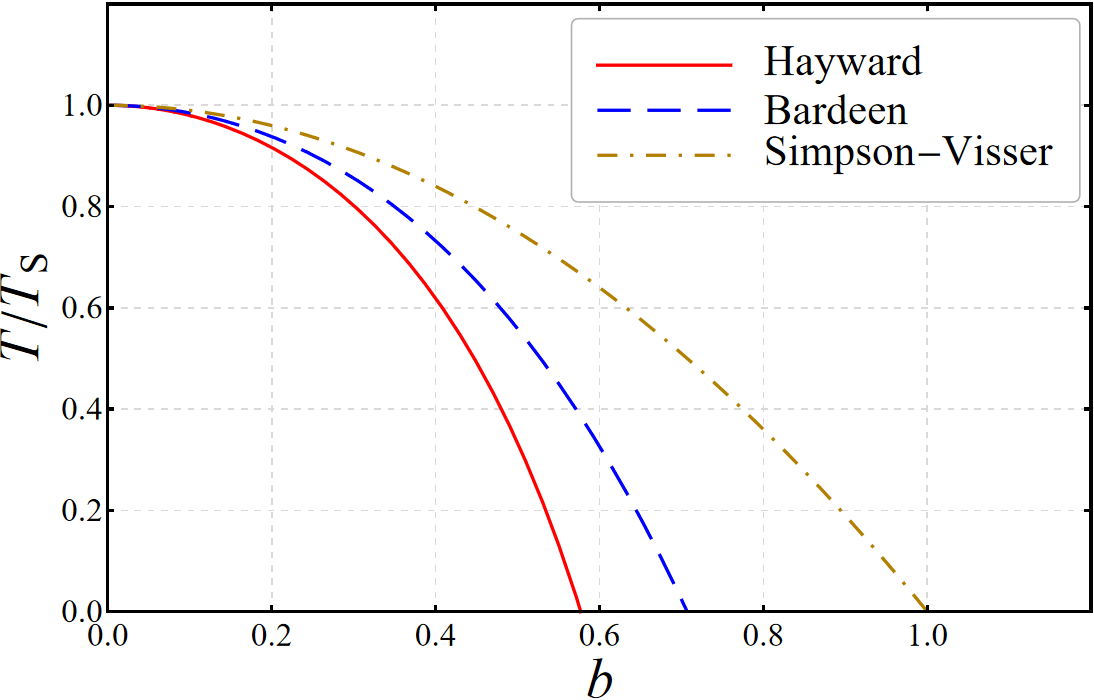}
\caption{The Hawking temperatures $T$ of three RBHs as a function of the parameter $b$. The red solid, blue dashed, and brown dot-dashed lines correspond to the Hayward, Bardeen, and Simpson--Visser black holes, with $b<1/\sqrt{3}$, $b<1/\sqrt{2}$, and $b<1$, respectively. The temperatures of all three RBHs decrease monotonically with $b$, with the Hayward case dropping most quickly. This observation indicates that the Hayward spacetime exerts the most pronounced influence on black hole evaporation.} \label{fig:Tn}
\end{figure}

To summarize, we provide the profiles of the black hole temperature (normalized by the Hawking temperature of the Schwarzschild black hole $T_{\rm S}$) as a function of the parameter $b$ in Fig. \ref{fig:Tn}. The specific models considered are the Hayward (red solid line), Bardeen (blue dashed line), and Simpson--Visser (brown dot-dashed line) black holes, respectively. All three lines exhibit similar decreasing behaviors, and the Hayward line drops most significantly, meaning that it has the most influential effect on black hole evaporation. It is evident that the constraints on the parameter $b$ from the positive definiteness of entropy are equivalent to the condition of a positive Hawking temperature. To highlight the deviations of RBHs from the Schwarzschild black hole, we will select $b=1/2$ for the subsequent analysis without loss of generality.

\section{MB effect of regular PBHs} \label{sec:dis}

Above, the corrections to the black hole temperature and entropy are computed for three types of RBHs. In this section, these results will be utilized to revise the evaporation dynamics of regular PBHs. We first derive the modified mass-loss rate and estimate the black hole lifetime, and then focus on the behavior of the MB-influenced PBHs during BBN. By constraining the impact of evaporation on light element synthesis, we discuss the viable mass windows for regular PBHs as a candidate of DM. 

\subsection{MB effect on the evaporation of regular PBHs} \label{sec:life}

We start from the model in which a regular PBH enters the MB phase after its half-decay. Afterwards, it continues to radiate at a suppressed emission rate until complete evaporation. The suppression factor is inversely proportional to the $k$-th power of the black hole entropy $S$. The case with $k=0$ corresponds to the semi-classical evaporation mode throughout the evolution, while the case with $k=1$ corresponds to the minimal suppression after half-decay.

\begin{figure}[htb]
\centering \includegraphics[width=0.7\linewidth]{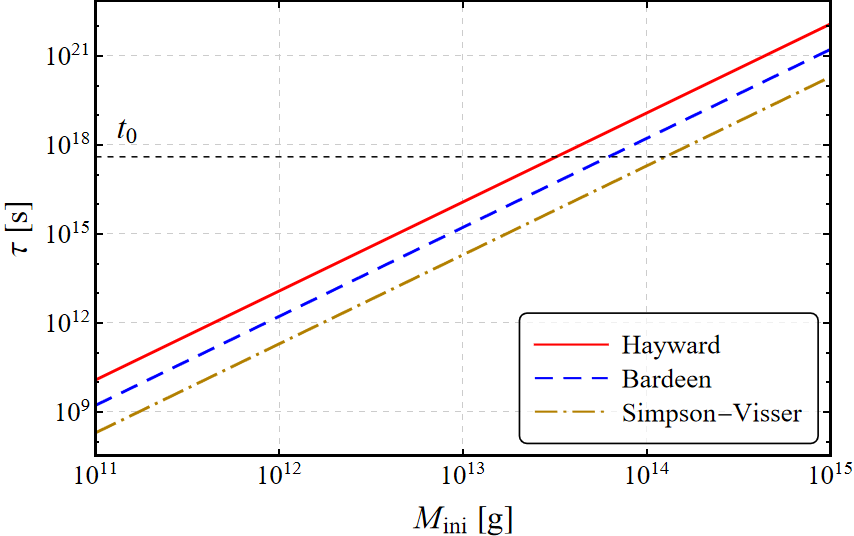}
\caption{The lifetimes $\tau$ as a function of initial mass $M_{\rm ini}$ for the Hayward (red solid line), Bardeen (blue dashed line), and Simpson--Visser (brown dot-dashed line) PBHs in the semi-classical evaporation, with the parameters $b=1/2$ and $k=0$. Here, $t_0$ represents the present age of the Universe. It is noteworthy that the lifetimes of these regular PBHs are longer than that of Schwarzschild black hole with the same mass, indicating that the regular metrics suppress the evaporation rate.} \label{fig:evaposc}
\end{figure}

First, during the semi-classical evaporation stage, we simulate the process by using the Stefan--Boltzmann law, meaning that we approximate the evaporation as a black-body spectrum, neglecting the gray-body factor.\footnote{The discussion on the impact of gray-body factor on the evaporation of RBHs can be found in Ref. \cite{Calza:2024fzo}, and for more specific calculations, we refer to Refs. \cite{Toshmatov:2015wga, MahdavianYekta:2019pol, Konoplya:2023ahd, Bolokhov:2024voa}. As it is not the primary focus of this work, we will not go into the details of this issue.} The evaporation equation is given by Eq. \eqref{SCevap}, and the lifetimes $\tau$ for the three types of regular PBHs are obtained as
\begin{align}
\tau_{\rr {H}}&= \frac{(1-b^2)^2}{(1-3b^2)^4} \tau_{\rm sc}, \label{tH}\\ 
\tau_{\rm{B}}&=\f{1+b^2}{3(1-2b^2)^4}\tau_{\rm sc}, \label{tB}\\ 
\tau_{\rr {SV}}&=\f{1}{3(1-b^2)^3}\tau_{\rm sc}. \label{tSV} 
\end{align}
It is straightforward to find that $\tau_{\rr {H}}$, $\tau_{\rr {B}}$, and $\tau_{\rr {SV}}$ are all proportional to $\tau_{\rm sc}$ (i.e., $M_{\rm ini}^3$), and $\tau_{\rr {H}}>\tau_{\rr {B}}>\tau_{\rr {SV}}>\tau_{\rr {sc}}$ when $b<1/\sqrt{3}$. This observation is consistent with Fig. \ref{fig:Tn}, as the black hole temperatures satisfy $T_{\rm H}<T_{\rm B}<T_{\rm SV}<T_{\rm S}$ with the same $b=1/2$. The relations between $\tau$ and the initial mass $M_{\mathrm{ini}}$ of regular PBHs are illustrated in Fig. \ref{fig:evaposc}, with $t_0= 4.35\times 10^{17}~\mathrm{s}$ being the age of the Universe.
It can be found that, in the semi-classical black hole evaporation, within the age of the Universe, the Hayward black hole with 
\begin{align}
M_{\rm{H}}> 3.2\times 10^{13}~\rm{g} \n 
\end{align}
can still survive today, whereas those with masses less than this value will have completely evaporated. For the Bardeen and Simpson--Visser black holes, these thresholds become 
\begin{align}
M_{\rm{B}} > 5.6\times 10^{13}~{\rm g}, \quad M_{\rm{SV}} > 1.3\times 10^{14}~\rm{g}. \n 
\end{align}
This observation indicates that, regular PBHs typically possess longer lifetimes than the Schwarzschild black hole with the same mass. 

\begin{figure}[htb]
\centering \includegraphics[width=0.7\linewidth]{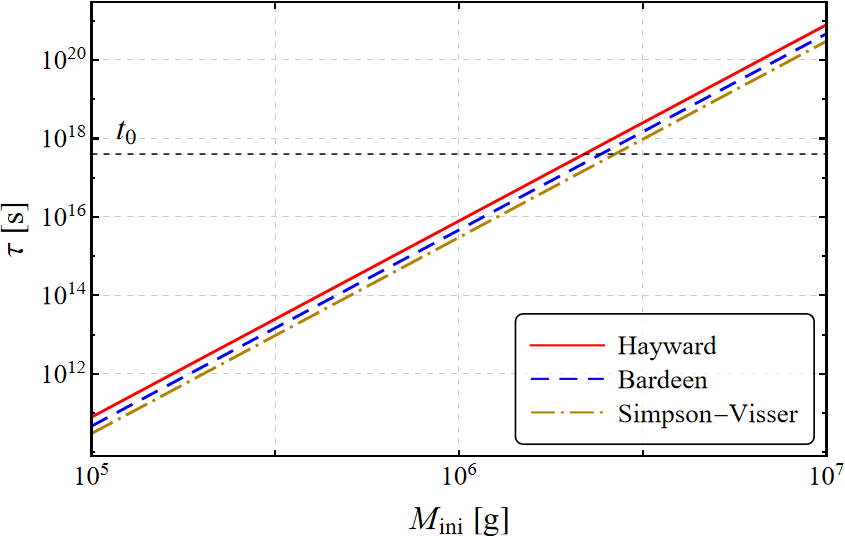}
\caption{Same as Fig. \ref{fig:evaposc}, but with the MB effect taken into account (i.e., $b=1/2$ but $k=1$ rather than 0). The discrepancies among the lines for the three types of regular PBHs become notably reduced. This indicates that the Hawking evaporation rates of these regular PBHs result in very similar behaviors, demonstrating that the MB effect plays a dominant role in the evaporation process.} \label{fig:evapomb}
\end{figure}

Now, we take into account the influence of the MB effect, and denote the black hole mass at the onset of the MB phase as $M^{(k)}$. When the black hole loses half of its initial mass $M_{\rm ini}$, it enters the MB phase. From a conservative point of view, we choose the MB parameter as $k=1$. From Eq. \eqref{MBF}, the black hole evaporation rate is thus given by
\begin{align}
\frac{\dd M^{(1)}}{\dd t} =\frac{1}{S}\frac{\dd M^{(0)}}{\dd t}. \label{evapomb}
\end{align}
The relations between the total lifetime $\tau$ and the initial mass $M_{\rm ini}$ of three types of regular PBHs, with the MB effect considered, are illustrated in Fig. \ref{fig:evapomb}. It can be seen that the black holes with
\begin{align}
M _{\rm{ini}} \gtrsim 2\times 10^{6}\,\rr{g} \label{lifetime}
\end{align}
are expected to survive up to now (almost the same for three regular PBHs). This modification implies a potential relaxation of the observational constraints, thereby opening a novel mass window within the parameter space that was previously excluded. Due to the significant impact of the MB effect, the lifetime scaling with respect to the initial mass of black hole entering this phase shifts from $\tau\propto M_{\rr{ini}}^3$ to $\tau\propto M_{\rr{ini}}^5$. Consequently, the MB effect tends to eliminate the discrepancies among different regular PBHs, so the influence of the specific RBH metric becomes negligible, and the lifetimes of the Hayward, Bardeen, and Simpson--Visser black holes become comparable with identical masses.

\subsection{BBN constraints} \label{sec:BBN}

To compare our theoretical results with experimental observations, we focus on the evaporation of regular PBHs during and after BBN, which increase the baryon-to-entropy ratio at the time of nucleosynthesis, resulting in an overabundance of $^4\rr {He}$ and a deficiency of deuterium. The emission of high-energy nucleons and anti-nucleons enhances the primordial deuterium abundance through the capture of free neutrons by protons and the spallation of $^4\rr {He}$. Furthermore, photons emitted by PBHs with $M>10^{10}~\rr g$ can dissociate the deuterons produced during nucleosynthesis. Given the remarkable successes of the standard BBN model in predicting light element abundances, any modifications are subject to strict constraints.

The PBHs with lifetimes $\tau$ shorter than $10^{-2}~\rr s$ are not constrained by BBN, as they have evaporated completely before weak freeze-out, without leaving any trace. The PBHs with $\tau\approx 10^{-2}$--$10^{4}~\rr s$ are constrained by the emission of mesons and anti-nucleons, which induce additional conversion between the protons and neutrons. This increases the freeze-out ratio of $n_n/n_p$ and the final abundance of $^4\rr{He}$. Finally, for the PBHs with $\tau \approx 10^2$--$10^7\,\rr s$, hadronic dissociation becomes significant, and the constraints on the PBH abundance $f_{\pbh}$ arise from the fragment deuterons and the non-thermal production of $^6\rr{Li}$ \cite{Carr:2020gox}.

Substituting Eqs. (\ref{tH})--(\ref{tSV}) into Eq. (\ref{ttMB}), we find that, if the three types of regular PBHs enter the MB-dominated phase prior to the onset of BBN ($t_{\rm BBN}\sim 1 \,\rr{s}$), their initial masses should lie in the ranges as
\begin{align}
M_{\rr {H}}<4.6\times 10^7\,\rr g, \quad
M_{\rr {B}}<8.8\times 10^7\,\rr g, \quad
M_{\rr {SV}}<1.8\times 10^8\,\rr g. \n
\end{align}
Consequently, their radiated energy will be suppressed. We expect that these regular PBHs have not evaporated completely and can survive to the present time as a candidate of DM. For broader mass ranges, the existing BBN constraints remain valid. Below, we will investigate the impact of such regular PBHs on BBN and thereby derive the constraints on their abundances.

Following the method outlined in Ref. \cite{Alexandre:2024nuo}, and assuming that the PBHs formed in the radiation-dominated era have a monochromatic mass distribution, we obtain the PBH number density as
\begin{align}
n_{\pbh}(T)\sim \frac{\rho_{\pbh}(t_0)}{M}\lt(\frac{T}{T_0}\rt)^3. \label{nPBH}
\end{align}
Here, we define a new parameter $\epsilon$ to quantify the ratio of the energy injected by PBHs with respect to the radiation energy associated with the Hubble expansion,
\begin{align}
\epsilon\equiv\lt|\frac{\dot{\rho}_{\pbh}}{\dot{\rho}_{\rm Hubble}}\rt|, \label{epsilon}
\end{align}
where $\rho_{\pbh}$ and $\rho_{\rm Hubble}$ are the energy densities of the PBHs and the radiation associated with the Hubble expansion. By calculating $\epsilon$ at different temperature $T$, we are able to determine whether the regular PBHs within certain mass range are constrained.

From Eq. (\ref{epsilon}), a direct calculation yields 
\begin{align}
\epsilon(t_{\rm BBN})\sim10^{-23}\ll 1. \n
\end{align} 
This indicates that the injected energy from PBHs is negligible compared to the radiation energy governed by the Hubble expansion, implying that any physical processes occurring during this period remain unaffected. A similar calculation applied to the CMB epoch yields
\begin{align}
\epsilon(t_{\rr{CMB}})\sim 10^{-5}. \n
\end{align}
Although this value is significantly larger than that in the BBN case, the condition $\epsilon(t_{\rr{CMB}})\ll 1$ remains satisfied. Thus, the regular PBHs entering the MB phase during this period appear to be unconstrained. However, the regular PBHs entering the MB phase at the CMB epoch ($t_{\rr{CMB}}\sim 10^{13}\,\rr s$) must have undergone the semi-classical evaporation during the BBN era, so they must be subject to the BBN constraints at earlier times. Since the BBN bounds are far more stringent than those derived from CMB, we only need to restrict our discussion to the viable parameter space for DM allowed by BBN.

Now, we explain in detail the method to constrain the regular PBH abundance $f_\pbh$ from BBN. We consider an extreme scenario, in which the regular PBHs exclusively radiate photons, and demand the radiated photon energy density to be less than the background photon energy density at the epoch of BBN, so we have
\begin{align}
n_{\pbh}\Delta E(t_{\rr{BBN}}) < \rho_{\gamma}(t_{\rr{BBN}}), \label{Ein}
\end{align}
where $\Delta E(t_{\rr{BBN}})$ is the energy radiated by a single PBH, and can be approximated as $\Delta E(t_{\rr {BBN}})=|\dot{M}|\,\Delta t$ (with $\Delta t \sim 10^3~\rr s$ being the duration of BBN), and $\rho _{\gamma}(t_{\rm BBN})$ is the background photon energy density, following the Stefan--Boltzmann law,
\begin{align}
\rho _{\gamma}(t_{\rm BBN}) =\frac{\pi ^2}{15}T_{\rr{BBN}}^4. \label{rhor}
\end{align}
Then, combing Eqs. (\ref{nPBH}), (\ref{Ein}), and (\ref{rhor}), we are able to obtain the PBH abundance $f_\pbh$ as
\begin{align}
f_{\pbh}(M)\equiv\frac{\rho _{\pbh}(t_0)}{\rho _{\rr {DM}}(t_0)}<\frac{\pi ^2}{15}\frac{T_{\rr {BBN}}T_0^3}{\rho_{\rr {DM}}(t_0)} \frac{M}{|\dot{M}|\,\Delta t}, \label{fPBH}
\end{align}
where $\rho_{\rm{DM}}(t_0)$ is the energy density of DM today. To make the constraint as stringent as possible, we select the parameter $T_{\rr {BBN}}$ as its minimum (around $30~\rr{keV}$), while the expression for $\dot{M}$ is collectively given by Eqs. \eqref{SCevap} and \eqref{evapomb}.

\begin{figure}[htb]
\centering \includegraphics[width=0.7\linewidth]{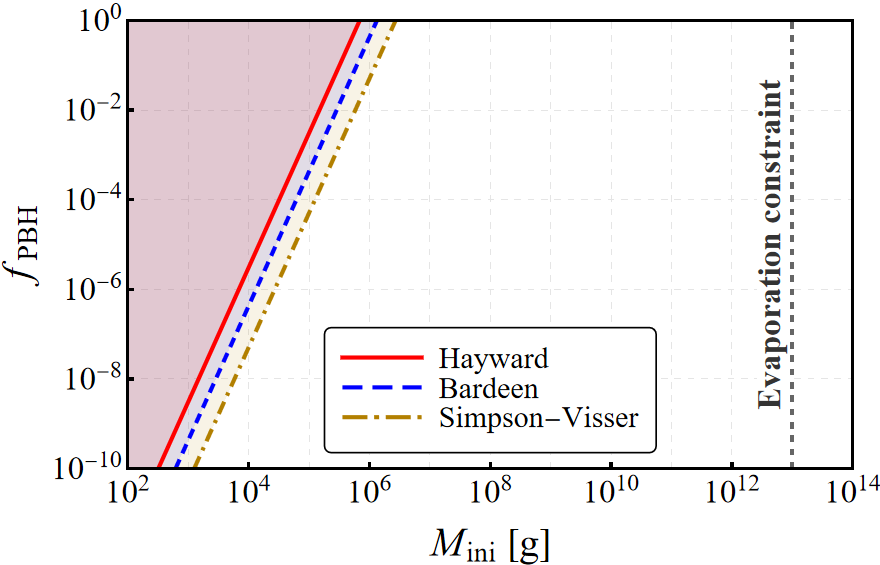}
\caption{The upper limits on the regular PBH abundance $f_{\pbh}$ as a function of the initial mass $M_{\rr{ini}}$, derived from the BBN constraints. The red solid, blue dashed, and brown dot-dashed lines correspond to the Hayward, Bardeen, and Simpson--Visser black holes, respectively. The colored regions above the relevant lines are excluded. The gray dashed line on the right denotes the constraints arising from black hole evaporation. It is evident that the constrained mass range is significantly smaller than the constraint imposed by evaporation, which implies that the mass windows for considering regular PBHs as a candidate of DM remain completely open.} \label{fig:fPBHsc}
\end{figure}

The bounds on the regular PBH abundance $f_{\pbh}$ as a function of its initial mass $M_{\rr{ini}}$, derived from the BBN constraints, are shown in Fig. \ref{fig:fPBHsc}. Only the regular PBHs with $M_{\rr{ini}}$ exceeding their thresholds in Eq. \eqref{lifetime} can survive to the present time and persist as DM. The lines on the left-hand-side place the upper limits on $f_\pbh$. Since the Hayward black hole receives the most significant correction on its temperature, its evaporation is suppressed to the greatest extent. Consequently, the BBN constraints on the Hayward black hole are the weakest. It is also worth noting that the parameter space constrained by BBN for these three types of regular PBHs is, in fact, smaller than that for the Schwarzschild black hole. As illustrated in Fig. \ref{fig:fPBHsc}, the BBN constraints on the regular PBH evaporation do not extend beyond $M_{\rm ini}\sim 10^7~\rr g$, which is significantly below the lower mass limit imposed by the evaporation constraint, $M_{\rm ini}\sim 10^{13}~\rr g$, which is determined by the evaporation of the Hayward black hole, representing the lowest mass limit among the three scenarios. 

\begin{figure}[htb]
\centering \includegraphics[width=0.7\linewidth]{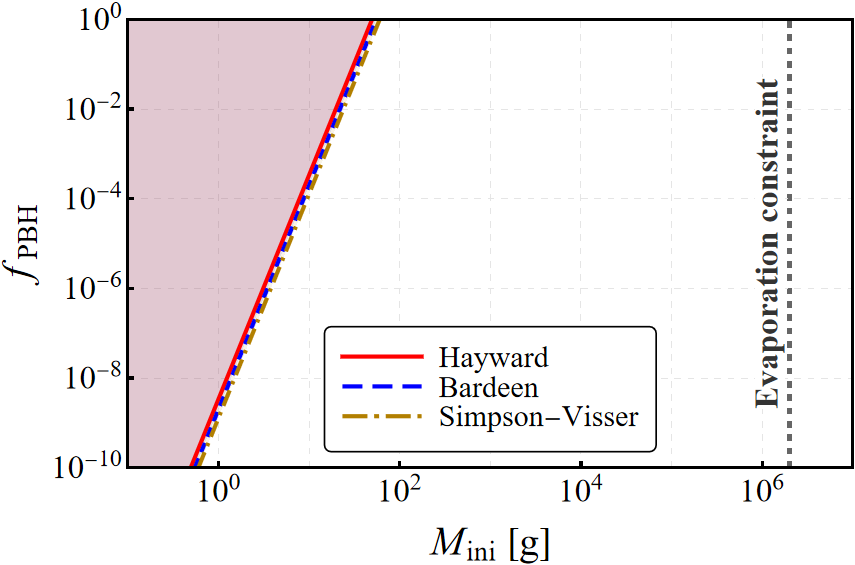}
\caption{Same as Fig. \ref{fig:fPBHsc}, but with the MB effect taken into account. The constrained mass range is $M_{\rm ini}\lesssim 10^2\,\rr g$, which is significantly smaller than the lower mass limit of $M_{\rm ini}\sim 2\times 10^6\,\rr g$ imposed by the evaporation constraint. Consequently, we can conclude that the regular PBHs serving as a candidate of DM have a negligible impact on BBN.} \label{fig:fPBHmb}
\end{figure}

Furthermore, the bounds on the regular PBH abundance $f_{\pbh}$ as a function of its initial mass $M_{\rm ini}$, derived from the BBN constraints, with the MB effect taken into account, are shown in Fig. \ref{fig:fPBHmb}. It can be observed that, for the regular PBHs entering the MB phase, the BBN constraint extends up to a maximum of $10^2~\rr g$, which is far below the mass window of PBHs capable of surviving to the present time. The upper limits on $f_\pbh$ for the three types of regular PBHs are nearly identical. This corroborates the conclusion drawn from the evaporation calculations: the MB effect is so dominant, that the influence of the specific regular spacetime metrics is almost irrelevant. Therefore, we can confidently assess that the following mass windows are effectively unconstrained from BBN and allow the regular PBHs to serve as total DM at the present time,
\begin{align}
2.2\times 10^6\,\rr g&< M_{\rr H}< 4.6 \times 10^7\,\rr g, \label{messH}\\
2.4\times 10^6\,\rr g&< M_{\rr B}< 8.8 \times 10^7\,\rr g, \label{messB}\\
2.7\times 10^6\,\rr g&< M_{\rr {SV}}< 1.8 \times 10^8\,\rr g. \label{messSV}
\end{align} 

In summary, we arrive at our basic conclusion: the non-singular metrics of regular PBHs can reduce the black hole evaporation rate, and the MB effect can further wash out the discrepancies among these regular PBHs. As a result, a new mass window for PBHs at around $10^6$--$10^8$~g is completely open, which was excluded previously.

\subsection{$k$-dependence of RBH evaporation} \label{sec:k}

Last, we investigate an important issue of the impact of the model parameter $k$ in the MB effect (see Eq. \eqref{MBF}) on the RBH evaporation. In the preceding discussions, we adopt $k=1$ as a benchmark to explore the phenomenological implications of the MB effect. However, the exact strength of the MB effect encoded in $k$ is inherently model-dependent.
Therefore, to provide a comprehensive picture and address the relevant parameter dependency, we will systematically study the dependence of the evaporation dynamics and observational constraints to the variation of $k$, with its range reasonably being $k\in [0.2,2.0]$.

\begin{figure}[htb]
\centering \includegraphics[width=0.7\linewidth]{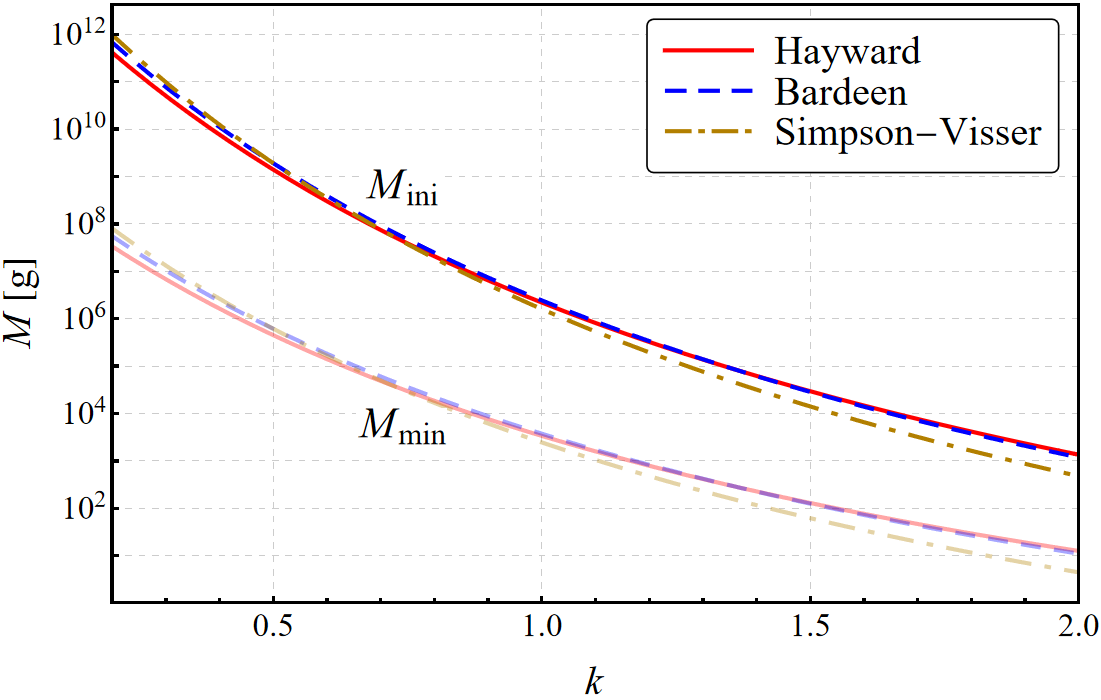}
\caption{The initial masses $M_{\mathrm{ini}}$ required to survive to the present day (upper dark curves) and the minimum masses $M_{\mathrm{min}}$ that evade the BBN constraints (lower light curves) for the Hayward, Bardeen, and Simpson--Visser black holes, as a function of the MB strength parameter $k$. The lower mass bounds imposed by evaporation consistently exceed those derived from the BBN constraints, ensuring that our preceding discussions remain valid.} \label{fig:k-M}
\end{figure}

As $k$ increases, the evaporation of regular PBHs is remarkably suppressed. From Eq. \eqref{MBF}, once a black hole enters the MB-dominated phase, the scaling relation between its lifetime and initial mass becomes $\tau \propto M_{\mathrm{ini}}^{2k+3}$. Hence, the lower mass bound imposed by the evaporation constraint is further reduced, thereby significantly broadening the allowed mass window for PBHs to constitute the entirety of DM, as illustrated in Fig. \ref{fig:k-M}. The minimum black hole mass $M_{\rm min}$ derived from the BBN constraints is also plotted for comparison.

An interesting phenomenon emerges when comparing the three RBH models across different $k$ values. As shown in Fig. \ref{fig:k-M}, the degree to which the MB effect influences different regular PBHs varies with $k$, leading to the intersections of the relevant curves.
For small values of $k<0.8$, the temperature decrease inherently induced by the metric geometry dominates the evaporation process, yielding the lowest mass bound for the Hayward black hole. However, as $k$ increases, the $S^{-k}$ suppression factor eventually takes over. As a result, the regular PBH with a relatively larger entropy will experience a more profound suppression of the evaporation rate. For instance, when $k> 1.3$, the Simpson--Visser black hole features the lowest mass bound among all the three models.

\begin{figure}[htb]
\centering
\begin{subfigure}
\centering\includegraphics[width=0.7\linewidth]{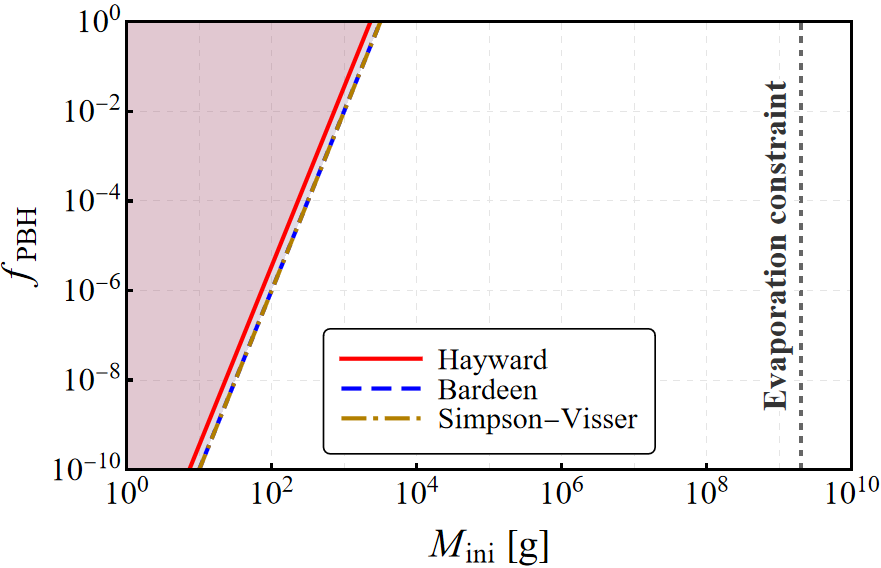}
\label{fig:k=0.5}
\end{subfigure}
\begin{subfigure}
\centering \includegraphics[width=0.7\linewidth]{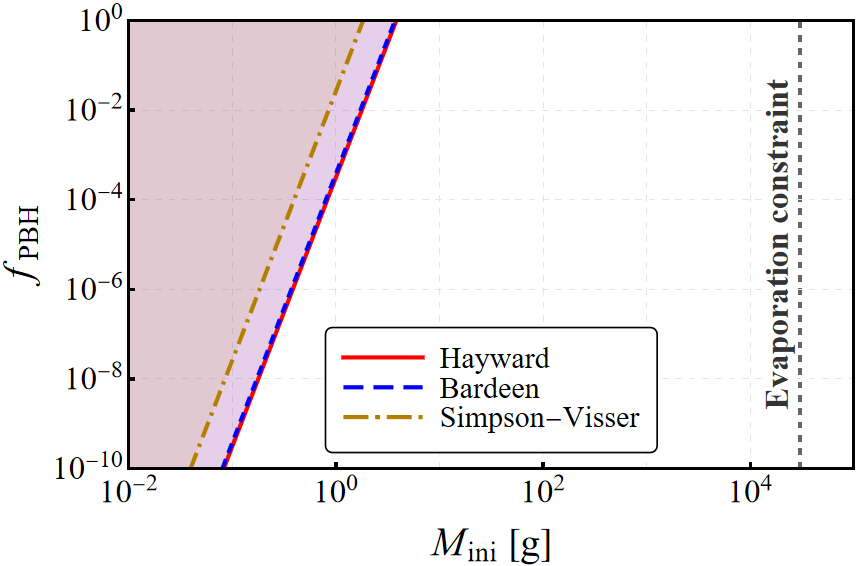}
\end{subfigure}
\caption{Same as Fig. \ref{fig:fPBHmb}, but with the MB strength parameter $k=0.5$ (upper panel) and $1.5$ (lower panel). Again, the mass ranges excluded by BBN remain strictly below the lower limits imposed by black hole evaporation. Consequently, the newly proposed mass windows remain fully open. From the three panels corresponding to $k=0.5$, $1$, and $1.5$, we observe a non-trivial interplay between the MB effect and the specific RBH metrics, as $k$ increases.} \label{fig:k=1.5}
\end{figure}

Although the lower limit imposed by the evaporation constraint decreases as $k$ increases, as qualitatively analyzed in Eq. \eqref{epsilon}, the lower bound imposed by BBN consistently remains far below that dictated by the evaporation lifetime across the entire range of $k \in [0.2, 2.0]$ in Fig. \ref{fig:k-M}, where $M_{\mathrm{ini}}$ is always larger than $M_{\mathrm{min}}$. The lower light curves exhibit a tendency identical to that of the upper dark curves, further demonstrating that the physical properties of all three RBH models are dependent on the intensity of the MB effect. Detailed BBN constraints of the PBH abundance for the specific cases of $k=0.5$ and $1.5$ are presented in Fig. \ref{fig:k=1.5}. Consequently, we conclude that, regardless of the varying strength of the MB effect, the mass window for regular PBHs to constitute all DM remains fully open towards the low-mass regime, and it is only the exact boundaries of this window 
that depend on the specific value of $k$.

\section{Conclusion} \label{sec:con}

PBHs have long been regarded as a promising non-particle candidate of DM. However, the Schwarz-schild PBHs with $M<10^{15}$~g have already evaporated before today. Therefore, to extend the allowed PBH mass range to even smaller values is becoming an interesting research area. In previous studies, two different directions have been explored in the literature. One is the RBHs, whose Hawking temperature is lower than that of the Schwarzschild case, thus reducing the radiation rate. The other is the MB effect, which breaks the assumption of self-similarity at the late stage of evaporation when the black hole has lost some of its initial mass.


Our work is a comprehensive study of the MB effect of three types of regular PBHs (the Hayward, Bardeen, and Simpson--Visser black holes, respectively). In this scenario, the evaporation rates are initially suppressed by the decrease of the Hawking temperature due to the spacetime regularization. After losing half of their initial masses, these regular PBHs enter the MB-dominated phase, where their evaporation rates are further attenuated by a suppression factor $S^{-k}$, so the discrepancies among them become even negligible. To bypass the issue of infinite-lifetime remnants,
we adopt a phenomenological self-similar ansatz for the evaporation process. Consequently, the MB effect significantly prolongs the PBH lifetimes, and thus drastically reduces the lower mass bounds, allowing us to reconsider the viability of light PBHs as a candidate of DM.

For the regular PBHs that can survive to the present time and have entered the MB-dominated phase prior to BBN, the evaporation rates are severely suppressed despite their small masses. This mitigates their potential impact on the BBN process, thereby relaxing the corresponding observational constraints. Our results indicate that, depending on the MB strength parameter $k$, regular PBHs can account for entire DM without violating the BBN bounds. For a benchmark scenario with $k=1$, this viable mass window is located at $10^6$--$10^8$~g (see Eqs. \eqref{messH}--\eqref{messSV}). Moreover, our systematic exploration reveals a non-trivial interplay between the MB effect and specific RBH metrics: as $k$ varies, the dominant suppression mechanism shifts, causing the favored RBH model to alternate. Altogether, while the exact mass boundaries are model-dependent, regular PBHs not only theoretically resolve the spacetime singularity problem, but also provide a phenomenologically promising framework for non-particle DM when combined with the MB effect.

\vskip .5cm

We are very grateful to Shi-Jie Wang, Qi-Lin Yang, and Yu-Jia Zhao for fruitful discussions. This work is supported by the National Natural Science Foundation of Liaoning Province, China (No. 2025-MS-034), and the National Training Program of Innovation and Entrepreneurship for Undergraduates (Project X202510145215).


{\small
\bibliography{refs}

}

\end{document}